\PassOptionsToPackage{unicode}{hyperref}
\PassOptionsToPackage{hyphens}{url}
\documentclass[
]{article}
\usepackage{xcolor}
\usepackage{amsmath,amssymb}
\usepackage{iftex}
\ifPDFTeX
  \usepackage[T1]{fontenc}
  \usepackage[utf8]{inputenc}
  \usepackage{textcomp} 
\else 
  \usepackage{unicode-math} 
  \defaultfontfeatures{Scale=MatchLowercase}
  \defaultfontfeatures[\rmfamily]{Ligatures=TeX,Scale=1}
\fi
\usepackage{lmodern}
\ifPDFTeX\else
\fi
\IfFileExists{upquote.sty}{\usepackage{upquote}}{}
\IfFileExists{microtype.sty}{
  \usepackage[]{microtype}
  \UseMicrotypeSet[protrusion]{basicmath} 
}{}
\makeatletter
\@ifundefined{KOMAClassName}{
  \IfFileExists{parskip.sty}{%
    \usepackage{parskip}
  }{
    \setlength{\parindent}{0pt}
    \setlength{\parskip}{6pt plus 2pt minus 1pt}}
}{
  \KOMAoptions{parskip=half}}
\makeatother
\usepackage{longtable,booktabs,array}
\usepackage{calc} 
\usepackage{etoolbox}
\makeatletter
\patchcmd\longtable{\par}{\if@noskipsec\mbox{}\fi\par}{}{}
\makeatother
\IfFileExists{footnotehyper.sty}{\usepackage{footnotehyper}}{\usepackage{footnote}}
\makesavenoteenv{longtable}
\usepackage{graphicx}
\usepackage{pgfplots}
\usepackage{pgfplotstable}
\pgfplotsset{compat=1.18}
\makeatletter
\newsavebox\pandoc@box
\newcommand*\pandocbounded[1]{
  \sbox\pandoc@box{#1}%
  \Gscale@div\@tempa{\textheight}{\dimexpr\ht\pandoc@box+\dp\pandoc@box\relax}%
  \Gscale@div\@tempb{\linewidth}{\wd\pandoc@box}%
  \ifdim\@tempb\p@<\@tempa\p@\let\@tempa\@tempb\fi
  \ifdim\@tempa\p@<\p@\scalebox{\@tempa}{\usebox\pandoc@box}%
  \else\usebox{\pandoc@box}%
  \fi%
}
\def\fps@figure{htbp}
\makeatother
\providecommand{\tightlist}{%
  \setlength{\itemsep}{0pt}\setlength{\parskip}{0pt}}
\usepackage{bookmark}
\IfFileExists{xurl.sty}{\usepackage{xurl}}{} 
\hypersetup{
  pdftitle={Verifier-First Evaluation of Agentic LLMs for Infrastructure-as-Code Generation},
  pdfauthor={Mohamed Jouini},
  hidelinks,
  pdfcreator={LaTeX via pandoc}}

\title{Verifier-First Evaluation of Agentic LLMs for
Infrastructure-as-Code Generation}
\author{Mohamed Jouini}
\date{}

\begin{document}
\maketitle

\section*{Abstract}\label{abstract}
\addcontentsline{toc}{section}{Abstract}

Infrastructure-as-Code (IaC) generation from natural language requires
satisfying provider schemas, dependency planning, and organizational
policy constraints --- not merely producing syntactically plausible
configurations. We present a verifier-first empirical study of seven
agentic strategies for Terraform generation evaluated on
\textbf{IaC-Eval v2}, a modernized 186-task AWS/Terraform benchmark with
Rego v1 intent policies. Our evaluation separates failures into three
verifier stages (\texttt{terraform\ validate}, \texttt{terraform\ plan},
\texttt{opa\ eval}) and applies McNemar's test with Wilson confidence
intervals on all pairwise comparisons (\(n=186\), \(\alpha=0.05\)).

We report five principal findings. \textbf{(1)} Active retrieval via
ReAct agents with MCP or ChromaDB-backed RAG raises Qwen2.5-Coder 7B
from 14.0\% to 45.7\% pass@1 (\(p<0.0001\)), primarily by reducing
VALIDATE\_FAIL from 144 to 66 tasks. \textbf{(2)} Iterative refinement
with verifier feedback achieves 62.9\% (Qwen 7B) and 84.4\% (GPT-4o)
pass@1, exhibiting binary convergence --- tasks either resolve in one
retry or exhaust the budget. \textbf{(3)} GEPA reflective instruction
optimization raises the Active RAG baseline by +7.5 pp (\(p=0.026\))
using only 80 verifier-guided rollouts, providing evidence that prompt
optimizers can improve verifiable IaC generation without weight updates.
\textbf{(4)} SIMBA teacher-free demonstration injection achieves
performance equivalent to Active RAG (\(p=1.0\)) without retrieval
infrastructure, but fails to address the dominant
SELF\_DEFINED\_PROPERTY error class (50\% of failures). \textbf{(5)} A
diagnostic Rego-injection experiment shows that 79\% of post-refinement
OPA failures are information-gap failures resolvable when policy text is
visible (\(p=0.016\)), motivating policy-aware prompt construction
rather than treating all policy failures as model-capability failures.

These results argue that IaC generation evaluation requires
stage-specific failure decomposition, that retrieval and repair target
orthogonal failure classes, and that instruction-level optimization
(GEPA) outperforms demonstration-level optimization (SIMBA) for
schema-constrained code generation tasks.

\textbf{Keywords:} Infrastructure-as-Code, Terraform, LLM code
generation, agentic AI, DSPy, ReAct, retrieval-augmented generation,
iterative refinement, prompt optimization, Open Policy Agent

\begin{center}\rule{0.5\linewidth}{0.5pt}\end{center}

\section{Introduction}\label{introduction}

Infrastructure-as-Code (IaC) enables cloud resources to be defined,
versioned, reviewed, and audited through declarative configuration files
{[}HashiCorp, 2024{]}. Terraform uses HashiCorp Configuration Language
(HCL) with a typed resource-block model spanning over 1,000 AWS resource
types. A correct Terraform configuration must simultaneously satisfy
four constraints: syntactic validity against the provider plugin,
schema-level conformance to the current provider version,
dependency-graph coherence across referenced resources, and compliance
with organizational policy specifications.

These layered requirements distinguish IaC generation from standard
code-completion benchmarks. Kon et al.~{[}2024{]} demonstrate that GPT-4
achieves 86.6\% on tightened Python tiers but only 19.36\% on IaC-Eval
Terraform tasks --- a gap attributable to provider-schema breadth,
declarative semantics, and external policy gates absent from
function-level code synthesis. Zhang et al.~{[}2025{]} show that
iterative feedback substantially improves CloudFormation deployability
but do not evaluate policy compliance. Neither study systematically
evaluates modern agentic techniques (ReAct tool use, RAG-based
retrieval, DSPy optimization) under a unified, multi-stage verification
harness.

This paper addresses this gap through a \textbf{verifier-first empirical
study} organized around four research questions:

\begin{itemize}
\tightlist
\item
  \textbf{RQ1 (Retrieval):} When does active retrieval through ReAct
  agents with MCP or RAG improve Terraform correctness over static
  context injection?
\item
  \textbf{RQ2 (Repair):} Which verifier-stage failures are recoverable
  through iterative refinement, and which persist across attempts?
\item
  \textbf{RQ3 (Optimization):} Can DSPy optimizers (SIMBA, GEPA) improve
  local-model IaC generation without weight updates or retrieval
  infrastructure?
\item
  \textbf{RQ4 (Failure Semantics):} Which residual errors are model
  capability failures, and which are information gaps in the
  prompt--policy interface?
\end{itemize}

Every strategy is evaluated through the same three-stage pipeline,
enabling direct attribution of improvements to specific failure modes.
All results are statistically validated with McNemar's test on 186
paired binary outcomes.

\subsection{Contributions}\label{contributions}

This work makes six contributions:

\begin{enumerate}
\def\labelenumi{\arabic{enumi}.}
\item
  \textbf{IaC-Eval v2 benchmark artifact.} A modernized 186-task
  Terraform/AWS dataset with AWS provider \textasciitilde\textgreater{}
  6.0, Rego v1 policies, corrected benchmark defects, and Hugging Face
  packaging (\texttt{iac-eval-v2/iac-eval-v2}).
\item
  \textbf{Verifier-stage evaluation methodology.} Explicit separation of
  failures into VALIDATE\_FAIL, PLAN\_FAIL, and OPA\_FAIL categories
  with per-gate attribution of strategy improvements.
\item
  \textbf{Comparative study of seven agentic strategies} across three
  models (GPT-4o, GPT-5.4, Qwen2.5-Coder 7B), demonstrating that
  retrieval, repair, and optimization address different verifier stages.
\item
  \textbf{Application of GEPA and SIMBA to IaC generation} with
  verifiable OPA rewards, extending DSPy optimization from NLP tasks to
  schema-constrained code generation.
\item
  \textbf{Failure taxonomy and error-class hierarchy.} We show that
  SELF\_DEFINED\_PROPERTY (schema hallucination) is model-scale-limited:
  SIMBA (50\%) → GEPA (23\%) → GPT-4o (9\%), establishing that
  instruction optimization partially compensates for parametric
  knowledge gaps.
\item
  \textbf{Reproducible observability methodology.} Every reported metric
  is traceable to a specific MLflow trace ID, experiment, and per-span
  execution record, enabling independent verification of all claims.
\end{enumerate}

\begin{center}\rule{0.5\linewidth}{0.5pt}\end{center}

\section{Related Work}\label{related-work}

\subsection{IaC Benchmarks and
Evaluation}\label{iac-benchmarks-and-evaluation}

\textbf{IaC-Eval} {[}Kon et al., 2024{]} introduced a Terraform/AWS
benchmark with OPA-based intent verification, establishing that IaC
generation is substantially harder than compact code-completion tasks.
The original study reports low pass@1 for GPT-4 on Terraform (19.36\%)
despite much stronger performance on Python tiers (86.6\%), motivating
IaC as a distinct evaluation domain. Our work builds on IaC-Eval by (i)
modernizing the dataset to current Terraform/AWS provider versions, (ii)
separating PLAN\_FAIL from VALIDATE\_FAIL, and (iii) evaluating seven
orchestration strategies under a common verifier-first harness.

\textbf{IaCGen} {[}Zhang et al., 2025{]} studies deployability-centric
CloudFormation generation and shows that iterative error feedback
substantially improves deployability after up to 25 regeneration
attempts. Our study complements IaCGen by focusing on Terraform,
distinguishing validation from planning from policy failures, and
evaluating policy compliance via OPA --- a dimension absent from
IaCGen's CloudFormation-centric evaluation. The SELF\_DEFINED\_PROPERTY
error class we identify corresponds to IaCGen's ``Invalid Attribute''
category, confirming that schema hallucination is cross-language (HCL
and YAML).

\textbf{EXP-Bench} {[}Kon et al., 2025{]} evaluates agents on
long-horizon research tasks and finds that structured task decomposition
predicts agent performance. Our three-stage pipeline (Analyze → Generate
→ Assemble) aligns with this insight: each stage has narrower scope than
monolithic generation.

\subsection{LLM Code Generation and
Self-Repair}\label{llm-code-generation-and-self-repair}

General code-generation benchmarks such as \textbf{HumanEval} {[}Chen et
al., 2021{]} emphasize compact programs with executable tests but do not
capture IaC-specific requirements: provider-version drift, declarative
dependency graphs, and external policy engines. \textbf{Self-Refine}
{[}Madaan et al., 2023{]} demonstrates that iterative refinement with
self-feedback can improve generated programs, while empirical analyses
of self-repair {[}Olausson et al., 2023{]} show that repair is not
uniformly reliable. Our results extend this line by decomposing repair
outcomes by verifier stage: validation and planning errors are largely
repairable, while OPA policy failures require information absent from
the natural-language prompt.

\textbf{CEDAR} {[}Nashid et al., 2023{]} demonstrates retrieval-based
prompt selection for code-related few-shot learning, finding that
selecting relevant code examples improves generation quality. Our SIMBA
experiment provides a complementary finding: DSPy's automated
demonstration selection via trajectory contrast achieves performance
equivalent to active retrieval (McNemar \(p=1.0\)) but fails to address
schema-constraint errors that require abstract instruction rules rather
than concrete examples.

\subsection{Retrieval-Augmented
Generation}\label{retrieval-augmented-generation}

\textbf{RAG} {[}Lewis et al., 2020{]} motivates external knowledge
sources when model parameters alone are insufficient.
\textbf{Adaptive-RAG} {[}Jeong et al., 2024{]} conditions retrieval
behavior on query complexity. In our study, Active MCP and Active RAG
instantiate adaptive retrieval for Terraform provider documentation: the
model requests schema information during generation via ReAct tool
calls. Our finding that active retrieval outperforms static retrieval by
+12.9 pp (\(p=0.004\)) confirms Adaptive-RAG's thesis for IaC: agentic,
query-driven retrieval is more effective than pre-fetched context for
multi-resource compositions.

\subsection{Prompt Optimization}\label{prompt-optimization}

\textbf{DSPy} {[}Khattab et al., 2024{]} treats instructions and
demonstrations as optimizable parameters. \textbf{MIPROv2} {[}Opsahl-Ong
et al., 2024{]} demonstrates joint instruction and demonstration
optimization. \textbf{GEPA} {[}Agrawal et al., 2026{]} extends the
optimizer family with reflective prompt evolution guided by a strong
teacher LM, achieving +6\% average over GRPO across six NLP tasks with
35× fewer rollouts. \textbf{SIMBA} uses trajectory contrast to inject
demonstrations and rules without a teacher model.

To our knowledge, this is among the first studies to evaluate GEPA and
SIMBA on IaC generation with deterministic verifiable rewards. We find
that instruction-level evolution (GEPA: +7.5 pp over Active RAG,
\(p=0.026\)) improves schema-constrained code generation, while
demonstration-level injection (SIMBA: -0.5 pp vs.~Active RAG, \(p=1.0\))
matches retrieval performance but does not improve over it. The
GEPA-vs-SIMBA difference trends in favor of GEPA but is borderline at
\(\alpha=0.05\) (\(p=0.059\)), so we report it as suggestive rather than
conclusive.

\subsection{Agentic Evaluation and
Observability}\label{agentic-evaluation-and-observability}

\textbf{ReAct} {[}Yao et al., 2023{]} interleaves reasoning and
tool-invocation actions in a Thought → Action → Observation loop. We
implement ReAct agents with both MCP-based structured tool access
{[}Anthropic et al., 2024{]} and dense retrieval over ChromaDB,
demonstrating that the retrieval channel affects which resources the
model can correctly generate but not the fundamental architecture of the
agent loop.

Our observability methodology --- MLflow experiment tracking with
per-span tracing of every LM call --- follows the growing concern that
agent evaluations should be traceable {[}Kon et al., 2025{]}. Every
percentage point we report can be expanded into task IDs, error classes,
and full execution traces.

\begin{center}\rule{0.5\linewidth}{0.5pt}\end{center}

\section{Benchmark and Methodology}\label{benchmark-and-methodology}

\subsection{IaC-Eval v2 Dataset}\label{iac-eval-v2-dataset}

IaC-Eval v2 is a 186-task Terraform/AWS benchmark derived from IaC-Eval
{[}Kon et al., 2024{]}, modernized along four dimensions: (i) Terraform
constraints updated to \(\geq\) 1.15 with AWS provider
\textasciitilde\textgreater{} 6.0; (ii) Rego policies migrated to Rego
v1 (\texttt{import\ rego.v1}); (iii) benchmark defects corrected (false
OPA failures from unknown-after-apply values, malformed Rego
expressions); (iv) all reference solutions revalidated. The dataset is
released as \texttt{iac-eval-v2/iac-eval-v2} on Hugging Face.

Each task comprises a natural-language prompt, a reference HCL solution,
and a Rego intent specification. Tasks span 186 AWS scenarios at six
difficulty levels, covering EC2, S3, Lambda, RDS, MSK, SageMaker, VPC,
IAM, Lightsail, DynamoDB, ElastiCache, and Lex service families.

\subsection{Evaluation Pipeline}\label{evaluation-pipeline}

Our harness implements a three-stage \textbf{verifier-first} pipeline:

\[
\text{HCL} \xrightarrow{\text{validate}} \text{PASS/FAIL} \xrightarrow{\text{plan}} \text{PASS/FAIL} \xrightarrow{\text{opa eval}} \text{PASS/FAIL}
\]

Failures are classified into a taxonomy: - \textbf{VALIDATE\_FAIL}:
Provider schema or syntax errors (e.g., unsupported argument, invalid
block type) - \textbf{PLAN\_FAIL}: Dependency resolution or reference
errors (e.g., undeclared resource, missing file) - \textbf{OPA\_FAIL}:
Policy intent violations (valid and plannable code that does not satisfy
the Rego specification)

\textbf{Primary metric:} pass@1 --- the fraction of tasks passing all
three gates on the first attempt. For iterative strategies,
passItr@\(k\) with budget \(k=4\) (initial + 3 retries).

\textbf{Statistical tests:} All pairwise comparisons use McNemar's test
on 186 paired binary outcomes (continuity-corrected \(\chi^2\)).
Confidence intervals are Wilson score intervals at 95\%.

\subsection{Models}\label{models}

\begin{longtable}[]{@{}
  >{\raggedright\arraybackslash}p{(\linewidth - 8\tabcolsep) * \real{0.3043}}
  >{\centering\arraybackslash}p{(\linewidth - 8\tabcolsep) * \real{0.2174}}
  >{\centering\arraybackslash}p{(\linewidth - 8\tabcolsep) * \real{0.2174}}
  >{\raggedright\arraybackslash}p{(\linewidth - 8\tabcolsep) * \real{0.1304}}
  >{\raggedright\arraybackslash}p{(\linewidth - 8\tabcolsep) * \real{0.1304}}@{}}
\toprule\noalign{}
\begin{minipage}[b]{\linewidth}\raggedright
Model
\end{minipage} & \begin{minipage}[b]{\linewidth}\centering
Parameters
\end{minipage} & \begin{minipage}[b]{\linewidth}\centering
Context
\end{minipage} & \begin{minipage}[b]{\linewidth}\raggedright
Deployment
\end{minipage} & \begin{minipage}[b]{\linewidth}\raggedright
Role
\end{minipage} \\
\midrule\noalign{}
\endhead
\bottomrule\noalign{}
\endlastfoot
GPT-4o & Undisclosed & 128K & OpenAI API via LiteLLM & Strong baseline;
repair + MCP \\
GPT-5.4 & Undisclosed & 128K & OpenAI API via LiteLLM & CoT baseline;
GEPA teacher \\
Qwen2.5-Coder 7B & 7B & 32K & vLLM (AWQ 4-bit, local GPU) & Local-model
stress test \\
\end{longtable}

All experiments use temperature=0.0 and max\_tokens=2048. The LiteLLM
proxy provides a unified OpenAI-compatible API across all models.

\subsection{Agentic Strategies}\label{agentic-strategies}

\begin{longtable}[]{@{}
  >{\raggedright\arraybackslash}p{(\linewidth - 4\tabcolsep) * \real{0.35}}
  >{\raggedright\arraybackslash}p{(\linewidth - 4\tabcolsep) * \real{0.45}}
  >{\raggedright\arraybackslash}p{(\linewidth - 4\tabcolsep) * \real{0.20}}@{}}
\toprule\noalign{}
\begin{minipage}[b]{\linewidth}\raggedright
Strategy
\end{minipage} & \begin{minipage}[b]{\linewidth}\raggedright
Description
\end{minipage} & \begin{minipage}[b]{\linewidth}\raggedright
Model(s)
\end{minipage} \\
\midrule\noalign{}
\endhead
\bottomrule\noalign{}
\endlastfoot
CoT baseline & Zero-shot chain-of-thought prompting & All three \\
ReAct + MCP & ReAct agent with MCP documentation tools & GPT-4o, Qwen 7B \\
ReAct + RAG & ReAct agent with dense ChromaDB retrieval tools & Qwen 7B \\
Static RAG & Single pre-fetch of provider docs before a three-stage CoT pipeline & Qwen 7B \\
Iterative Refinement & Initial generation followed by up to three verifier-guided retries (\(k=4\)) & GPT-4o, Qwen 7B \\
Rego Oracle & Iterative refinement with full Rego policy text injected into the prompt & GPT-4o \\
GEPA-RAG & GEPA-optimized ReAct+RAG pipeline (reflection LM: GPT-5.4) & Qwen 7B \\
SIMBA & SIMBA-optimized passive-context pipeline, no retrieval at inference & Qwen 7B \\
\end{longtable}

\subsection{DSPy Optimization
Configuration}\label{dspy-optimization-configuration}

\textbf{GEPA} is applied to the ReAct+RAG pipeline with GPT-5.4 as the reflection LM. The optimizer runs for 80 verifier-guided rollouts with a 20-task Pareto validation set and a 124-task training set.

\textbf{SIMBA} is applied to the passive-context pipeline with no retrieval at inference time. The student model is Qwen~7B (teacher-free). The optimizer is run at a minimal configuration to establish a lower-bound result, with a 124-task training set.

\subsection{Observability and
Reproducibility}\label{observability-and-reproducibility}

Every experiment runs under a dual-sink observability stack. MLflow~v3 provides persistent per-span tracing of every LM call, module invocation, and tool use, with one experiment per strategy and one trace per task. Arize Phoenix provides real-time trace visualization via the OpenInference protocol.

The observability infrastructure is deployed on Kubernetes via Terraform. A command-line trace query toolkit provides programmatic access to all traces via the MLflow REST API, enabling independent verification of any reported result.

\begin{center}\rule{0.5\linewidth}{0.5pt}\end{center}

\section{Results}\label{results}

\subsection{Baseline Performance (RQ1--RQ4
Context)}\label{baseline-performance-rq1rq4-context}

Table 1 reports baseline CoT performance establishing the need for
staged evaluation.

\textbf{Table 1.} Baseline CoT performance (zero-shot, no retrieval, no
repair).

\begin{longtable}[]{@{}lcccccc@{}}
\toprule\noalign{}
Model & pass@1 & 95\% CI & valid@1 & VALIDATE & PLAN & OPA \\
\midrule\noalign{}
\endhead
\bottomrule\noalign{}
\endlastfoot
GPT-4o & 41.4\% & {[}34.5, 48.6{]} & 59.1\% & 76 & 15 & 18 \\
GPT-5.4 & 62.4\% & {[}55.2, 69.1{]} & 84.9\% & 28 & 14 & 28 \\
Qwen 7B & 14.0\% & {[}9.7, 19.7{]} & 22.6\% & 144 & 7 & 9 \\
\end{longtable}

\textbf{Finding 1.} Qwen 7B fails primarily at validation (144/186,
77.4\%), indicating that the dominant bottleneck for small models is
provider-schema hallucination. GPT-5.4's failures shift to OPA (28
tasks), demonstrating that a stronger model reaches later verifier
stages more frequently --- a phenomenon we term \textbf{failure
promotion}.

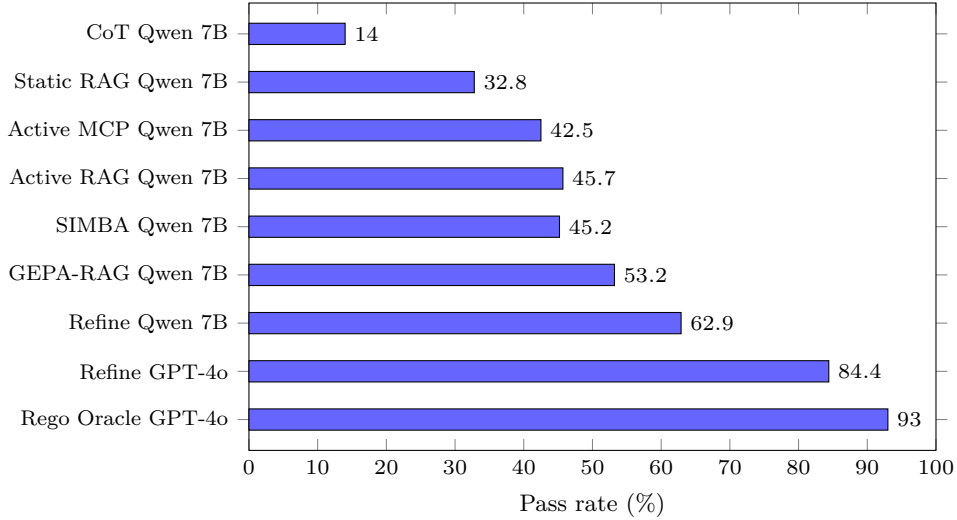
\begin{figure}[htbp]
\centering
\begin{tikzpicture}
\begin{axis}[
  xbar, xmin=0, xmax=100,
  width=0.88\linewidth, height=7.5cm,
  bar width=8pt,
  xlabel={Pass rate (\%)},
  symbolic y coords={
    {CoT Qwen 7B},
    {Static RAG Qwen 7B},
    {Active MCP Qwen 7B},
    {Active RAG Qwen 7B},
    {SIMBA Qwen 7B},
    {GEPA-RAG Qwen 7B},
    {Refine Qwen 7B},
    {Refine GPT-4o},
    {Rego Oracle GPT-4o}
  },
  ytick=data,
  y dir=reverse,
  nodes near coords, nodes near coords align={horizontal},
  every node near coord/.append style={font=\footnotesize},
  enlarge y limits=0.08,
  legend pos=south east,
  tick label style={font=\footnotesize},
  label style={font=\small},
]
\addplot[fill=blue!60] coordinates {
  (14.0,{CoT Qwen 7B})
  (32.8,{Static RAG Qwen 7B})
  (42.5,{Active MCP Qwen 7B})
  (45.7,{Active RAG Qwen 7B})
  (45.2,{SIMBA Qwen 7B})
  (53.2,{GEPA-RAG Qwen 7B})
  (62.9,{Refine Qwen 7B})
  (84.4,{Refine GPT-4o})
  (93.0,{Rego Oracle GPT-4o})
};
\end{axis}
\end{tikzpicture}
\caption{Overall pass@1 rate across all evaluated strategies (IaC-Eval v2, $n=186$). The performance ladder progresses from CoT through active retrieval, prompt optimization, iterative refinement, and the diagnostic policy oracle.}
\label{fig:leaderboard}
\end{figure}

\textbf{Figure 1.} Overall pass rate across evaluated strategies. The
main performance ladder is CoT \textless{} static retrieval \textless{}
active retrieval \textless{} prompt optimization \textless{} iterative
refinement \textless{} diagnostic policy oracle.

\subsection{RQ1: Active Retrieval Reduces Schema
Failures}\label{rq1-active-retrieval-reduces-schema-failures}

\textbf{Table 2.} Retrieval strategy comparison (Qwen2.5-Coder 7B, 186
tasks).

\begin{longtable}[]{@{}
  >{\raggedright\arraybackslash}p{(\linewidth - 14\tabcolsep) * \real{0.0789}}
  >{\centering\arraybackslash}p{(\linewidth - 14\tabcolsep) * \real{0.1316}}
  >{\centering\arraybackslash}p{(\linewidth - 14\tabcolsep) * \real{0.1316}}
  >{\centering\arraybackslash}p{(\linewidth - 14\tabcolsep) * \real{0.1316}}
  >{\centering\arraybackslash}p{(\linewidth - 14\tabcolsep) * \real{0.1316}}
  >{\centering\arraybackslash}p{(\linewidth - 14\tabcolsep) * \real{0.1316}}
  >{\centering\arraybackslash}p{(\linewidth - 14\tabcolsep) * \real{0.1316}}
  >{\centering\arraybackslash}p{(\linewidth - 14\tabcolsep) * \real{0.1316}}@{}}
\toprule\noalign{}
\begin{minipage}[b]{\linewidth}\raggedright
Strategy
\end{minipage} & \begin{minipage}[b]{\linewidth}\centering
pass@1
\end{minipage} & \begin{minipage}[b]{\linewidth}\centering
95\% CI
\end{minipage} & \begin{minipage}[b]{\linewidth}\centering
valid@1
\end{minipage} & \begin{minipage}[b]{\linewidth}\centering
VAL
\end{minipage} & \begin{minipage}[b]{\linewidth}\centering
PLN
\end{minipage} & \begin{minipage}[b]{\linewidth}\centering
OPA
\end{minipage} & \begin{minipage}[b]{\linewidth}\centering
Avg time
\end{minipage} \\
\midrule\noalign{}
\endhead
\bottomrule\noalign{}
\endlastfoot
CoT (no retrieval) & 14.0\% & {[}9.7, 19.7{]} & 22.6\% & 144 & 7 & 9 &
42s \\
Static RAG & 32.8\% & {[}26.5, 39.8{]} & 47.3\% & 98 & 15 & 12 & 54s \\
Active MCP (ReAct) & 42.5\% & {[}35.6, 49.7{]} & 55.9\% & 82 & 10 & 15 &
38s \\
Active RAG (ReAct) & 45.7\% & {[}38.7, 52.9{]} & 64.5\% & 66 & 14 & 21 &
36s \\
\end{longtable}

\textbf{Table 3.} Pairwise McNemar tests --- retrieval strategies
(\(n=186\)).

\begin{longtable}[]{@{}lcccl@{}}
\toprule\noalign{}
Comparison & Wins A & Wins B & \(p\)-value & Verdict \\
\midrule\noalign{}
\endhead
\bottomrule\noalign{}
\endlastfoot
ActiveRAG vs.~CoT & 67 & 8 & \(<0.0001\) & Significant \\
ActiveRAG vs.~StaticRAG & 43 & 19 & 0.004 & Significant \\
ActiveRAG vs.~MCP & 27 & 21 & 0.47 & Not significant \\
MCP vs.~StaticRAG & 35 & 17 & 0.018 & Significant \\
\end{longtable}

\textbf{Finding 2.} Active retrieval significantly outperforms static
retrieval (McNemar \(p=0.004\)). The +12.9 pp gap between Active RAG and
Static RAG is concentrated at the validation stage: VALIDATE\_FAIL drops
from 98 to 66 ($-$33\%). Active retrieval discovers dependencies and
nested provider schema details that keyword-based pre-fetching misses.

\textbf{Finding 3.} Active RAG and Active MCP are statistically
equivalent (\(p=0.38\)). ChromaDB dense retrieval edges MCP by +3.2 pp
in valid@1 (64.5\% vs.~55.9\%), attributable to pre-processed chunks
optimized for the 32K context window versus verbose MCP tool responses.

\textbf{Finding 4 (GPT-4o retrieval lift).} Active MCP raises GPT-4o
from 41.4\% to 70.4\% (+29.0 pp), reducing OPA\_FAIL to only 9 tasks
(4.8\%). For a strong model, retrieval largely addresses the schema
knowledge gap --- residual failures shift to policy compliance.

\subsection{RQ2: Iterative Refinement Recovers Validation
Failures}\label{rq2-iterative-refinement-recovers-validation-failures}

\textbf{Table 4.} Iterative refinement results (budget \(k=4\): initial
+ 3 retries).

\begin{longtable}[]{@{}
  >{\raggedright\arraybackslash}p{(\linewidth - 16\tabcolsep) * \real{0.0732}}
  >{\raggedright\arraybackslash}p{(\linewidth - 16\tabcolsep) * \real{0.0732}}
  >{\centering\arraybackslash}p{(\linewidth - 16\tabcolsep) * \real{0.1220}}
  >{\centering\arraybackslash}p{(\linewidth - 16\tabcolsep) * \real{0.1220}}
  >{\centering\arraybackslash}p{(\linewidth - 16\tabcolsep) * \real{0.1220}}
  >{\centering\arraybackslash}p{(\linewidth - 16\tabcolsep) * \real{0.1220}}
  >{\centering\arraybackslash}p{(\linewidth - 16\tabcolsep) * \real{0.1220}}
  >{\centering\arraybackslash}p{(\linewidth - 16\tabcolsep) * \real{0.1220}}
  >{\centering\arraybackslash}p{(\linewidth - 16\tabcolsep) * \real{0.1220}}@{}}
\toprule\noalign{}
\begin{minipage}[b]{\linewidth}\raggedright
Strategy
\end{minipage} & \begin{minipage}[b]{\linewidth}\raggedright
Model
\end{minipage} & \begin{minipage}[b]{\linewidth}\centering
pass@1
\end{minipage} & \begin{minipage}[b]{\linewidth}\centering
95\% CI
\end{minipage} & \begin{minipage}[b]{\linewidth}\centering
VAL
\end{minipage} & \begin{minipage}[b]{\linewidth}\centering
PLN
\end{minipage} & \begin{minipage}[b]{\linewidth}\centering
OPA
\end{minipage} & \begin{minipage}[b]{\linewidth}\centering
Avg time
\end{minipage} & \begin{minipage}[b]{\linewidth}\centering
Avg att.
\end{minipage} \\
\midrule\noalign{}
\endhead
\bottomrule\noalign{}
\endlastfoot
Active RAG (single-pass) & Qwen 7B & 45.7\% & {[}38.7, 52.9{]} & 66 & 14
& 21 & 36s & 1.0 \\
+ Repair (\(k=4\)) & Qwen 7B & 62.9\% & {[}55.8, 69.5{]} & 43 & 9 & 17 &
203s & 2.35 \\
Active MCP (single-pass) & GPT-4o & 70.4\% & {[}63.4, 76.6{]} & 27 & 19
& 9 & 60s & 1.0 \\
+ Repair (\(k=4\)) & GPT-4o & 84.4\% & {[}78.4, 89.1{]} & 8 & 7 & 14 &
98s & 1.3 \\
+ Rego Oracle (\(k=4\)) & GPT-4o & 93.0\% & {[}88.4, 96.0{]} & 1 & 9 & 3
& 86s & 1.2 \\
\end{longtable}

\textbf{Finding 5 (Binary convergence).} The repair-attempt distribution
is bimodal: 83\% of GPT-4o tasks pass in 1--2 attempts; 16\% exhaust the
full budget. Tasks either resolve immediately upon seeing the error or
do not resolve at all within \(k=4\). This validates the binary
convergence pattern first documented for general code self-repair
{[}Olausson et al., 2023{]} and extends it to IaC with verifier
feedback.

\textbf{Finding 6 (Repair promotes failures).} GPT-4o repair reduces
VALIDATE\_FAIL from 27→8 and PLAN\_FAIL from 19→7, but OPA\_FAIL
increases from 9→14. This is not regression --- it is failure promotion:
previously invalid configurations that now pass validation and planning
are exposed to the policy gate for the first time. IaCGen {[}Zhang et
al., 2025{]} reports a similar pattern for CloudFormation deployability
but does not evaluate policy compliance to observe this effect.

\textbf{Finding 7 (SELF\_DEFINED\_PROPERTY persistence under repair).}
For Qwen 7B, SELF\_DEFINED\_PROPERTY remains the largest manually coded
residual error class after refinement (27 of 69 failures, 39.1\%). The
repair loop cannot reliably correct schema hallucination when the
model's parametric knowledge does not contain the correct attribute name
--- the error message states ``argument not expected here'' but does not
suggest the correct alternative. This is an inherent limitation of
repair: it provides \emph{negative feedback} (what is wrong) but not
\emph{positive knowledge} (what is right). IaCGen {[}Zhang et al.,
2025{]} identifies ``Invalid Attribute'' as the analogous dominant
failure class for CloudFormation.

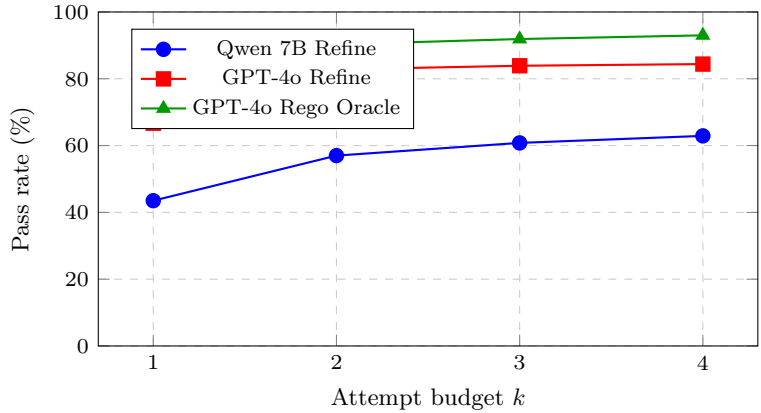
\begin{figure}[htbp]
\centering
\begin{tikzpicture}
\begin{axis}[
  width=0.85\linewidth, height=6cm,
  xlabel={Attempt budget $k$},
  ylabel={Pass rate (\%)},
  xtick={1,2,3,4},
  ymin=0, ymax=100,
  grid=major, grid style={dashed,gray!40},
  legend style={at={(0.05,0.95)},anchor=north west,font=\footnotesize},
  tick label style={font=\footnotesize},
  label style={font=\small},
  mark size=2.5pt,
]
\addplot[blue,mark=*,thick] coordinates {(1,43.5)(2,57.0)(3,60.8)(4,62.9)};
\addlegendentry{Qwen 7B Refine}
\addplot[red,mark=square*,thick] coordinates {(1,66.7)(2,82.8)(3,83.9)(4,84.4)};
\addlegendentry{GPT-4o Refine}
\addplot[green!60!black,mark=triangle*,thick] coordinates {(1,76.3)(2,90.3)(3,91.9)(4,93.0)};
\addlegendentry{GPT-4o Rego Oracle}
\end{axis}
\end{tikzpicture}
\caption{Iterative refinement convergence curves. GPT-4o achieves most gains by the second attempt; Qwen~7B benefits from additional retries but remains bounded by schema-knowledge failures.}
\label{fig:convergence}
\end{figure}

\textbf{Figure 2.} Refinement convergence by attempt budget. GPT-4o
obtains almost all gains by the second attempt, while Qwen 7B benefits
from additional retries but remains bounded by schema-knowledge
failures.

\subsection{RQ3: DSPy Optimization}\label{rq3-dspy-optimization}

\textbf{Table 5.} DSPy optimization results (Qwen2.5-Coder 7B, 186
tasks, single-pass).

\begin{longtable}[]{@{}
  >{\raggedright\arraybackslash}p{(\linewidth - 14\tabcolsep) * \real{0.0789}}
  >{\centering\arraybackslash}p{(\linewidth - 14\tabcolsep) * \real{0.1316}}
  >{\centering\arraybackslash}p{(\linewidth - 14\tabcolsep) * \real{0.1316}}
  >{\centering\arraybackslash}p{(\linewidth - 14\tabcolsep) * \real{0.1316}}
  >{\centering\arraybackslash}p{(\linewidth - 14\tabcolsep) * \real{0.1316}}
  >{\centering\arraybackslash}p{(\linewidth - 14\tabcolsep) * \real{0.1316}}
  >{\centering\arraybackslash}p{(\linewidth - 14\tabcolsep) * \real{0.1316}}
  >{\centering\arraybackslash}p{(\linewidth - 14\tabcolsep) * \real{0.1316}}@{}}
\toprule\noalign{}
\begin{minipage}[b]{\linewidth}\raggedright
Strategy
\end{minipage} & \begin{minipage}[b]{\linewidth}\centering
pass@1
\end{minipage} & \begin{minipage}[b]{\linewidth}\centering
95\% CI
\end{minipage} & \begin{minipage}[b]{\linewidth}\centering
VAL
\end{minipage} & \begin{minipage}[b]{\linewidth}\centering
PLN
\end{minipage} & \begin{minipage}[b]{\linewidth}\centering
OPA
\end{minipage} & \begin{minipage}[b]{\linewidth}\centering
Avg time
\end{minipage} & \begin{minipage}[b]{\linewidth}\centering
Teacher
\end{minipage} \\
\midrule\noalign{}
\endhead
\bottomrule\noalign{}
\endlastfoot
Active RAG (baseline) & 45.7\% & {[}38.7, 52.9{]} & 66 & 14 & 21 & 36s &
--- \\
SIMBA (passive ctx.) & 45.2\% & {[}38.2, 52.3{]} & 78 & 6 & 18 & 48s &
None \\
GEPA-RAG & 53.2\% & {[}46.1, 60.3{]} & 57 & 11 & 19 & 40s & GPT-5.4 \\
\end{longtable}

\textbf{Table 6.} Pairwise McNemar tests --- optimization strategies
(\(n=186\)).

\begin{longtable}[]{@{}lcccl@{}}
\toprule\noalign{}
Comparison & Wins A & Wins B & \(p\)-value & Verdict \\
\midrule\noalign{}
\endhead
\bottomrule\noalign{}
\endlastfoot
GEPA vs.~ActiveRAG & 24 & 10 & 0.026 & Significant \\
GEPA vs.~SIMBA & 35 & 20 & 0.059 & Borderline \\
SIMBA vs.~ActiveRAG & 29 & 30 & 1.00 & Not significant \\
SIMBA vs.~CoT & 63 & 5 & \(<0.0001\) & Significant \\
SIMBA vs.~StaticRAG & 41 & 18 & 0.004 & Significant \\
\end{longtable}

\textbf{Finding 8 (GEPA for IaC optimization).} GEPA instruction
optimization raises the Active RAG agent from 45.7\% to 53.2\% (+7.5 pp,
McNemar \(p=0.026\)) using only 80 verifier-guided rollouts with GPT-5.4
as reflection LM. This extends GEPA {[}Agrawal et al., 2026{]} to IaC
generation with OPA-gated verifiable rewards --- a qualitatively
different domain from the NLP tasks (HotpotQA, AIME, IFBench) in the
original GEPA evaluation. The +16.4\% relative improvement is consistent
with the upper range of GEPA's reported gains (+6\% average over GRPO
across six tasks).

\textbf{Finding 9 (GEPA automates schema-rule engineering).} Inspection
of the optimized instruction text reveals three domain-specific rules
learned from training failures: - \texttt{aws\_elasticache\_user}:
passwords must be 16--128 characters - \texttt{aws\_elasticache\_user}:
\texttt{engine\ =\ "redis"} (exact lowercase) -
\texttt{aws\_dynamodb\_contributor\_insights}: requires both table and
insights resources

These rules correspond to recurring failure patterns that GPT-5.4
identified by reflecting on verifier error messages. This is
\textbf{automated system-prompt engineering} --- GEPA discovers what
IaCGen {[}Zhang et al., 2025{]} engineers manually for CloudFormation.

\textbf{Finding 10 (SIMBA: teacher-free optimization matches
retrieval).} SIMBA achieves 45.2\% without retrieval infrastructure ---
statistically equivalent to Active RAG (45.7\%, \(p=1.0\)). This
demonstrates that teacher-free demonstration injection can match
architecturally more complex retrieval pipelines, requiring neither a
vector store nor a teacher model. However, SIMBA's single easy
demonstration (Task 265: \texttt{aws\_iam\_group}) provides no schema
guidance for complex resources, leaving SELF\_DEFINED\_PROPERTY
unchanged at 50\% of failures.

\textbf{Finding 11 (Instruction evolution is the stronger signal in this
setting).} GEPA (+7.5 pp over Active RAG, \(p=0.026\)) improves the
retrieval baseline, whereas SIMBA is statistically indistinguishable
from Active RAG (\(p=1.0\)). GEPA also trends above SIMBA (+8.0 pp), but
the paired comparison is borderline (\(p=0.059\)) and should not be
interpreted as a definitive superiority claim. The mechanistic
explanation is nevertheless informative: instructions encode
\emph{abstract rules} that generalize across task complexities (``only
use documented attributes''), while demonstrations anchor \emph{specific
output formats} that generalize mostly to structurally similar tasks.

\textbf{Finding 12 (SIMBA + GEPA complementarity).} SIMBA uniquely
solves 20 tasks GEPA misses; GEPA uniquely solves 35 tasks SIMBA misses.
Their union yields an oracle pass rate of 119/186 (64.0\%) --- nearly
identical to the observed iterative refinement result (62.9\%). This
suggests that: (a) single-pass optimization strategies collectively
approach the same bound as test-time adaptive correction; (b) an
ensemble or sequential application (SIMBA → GEPA) could achieve this
bound in a single pass.

\subsection{RQ4: Information-Gap vs.~Capability-Gap
Failures}\label{rq4-information-gap-vs.-capability-gap-failures}

\textbf{Table 7.} Rego oracle experiment --- GPT-4o repair → Rego
injection transitions.

\begin{longtable}[]{@{}lcl@{}}
\toprule\noalign{}
Transition & Tasks & Interpretation \\
\midrule\noalign{}
\endhead
\bottomrule\noalign{}
\endlastfoot
PASS → PASS & 156 & Already correct \\
OPA → PASS & \textbf{11} & Information gap (policy visible →
resolved) \\
VAL → PASS & 6 & Upstream fix from Rego context \\
OPA → OPA & \textbf{3} & Capability gap (model fails even with
policy) \\
PLAN → PLAN & 7 & Lambda sandbox cluster (structural) \\
\end{longtable}

\textbf{Finding 13 (79\% of OPA failures are information gaps).} Of 14
post-repair OPA failures, 11 resolve when the Rego policy text is
injected into the prompt. These tasks share a common structure: the
policy checks specific resource labels (e.g.,
\texttt{bucket.name\ ==\ "example"}) never communicated in the
natural-language prompt. The model generates functionally correct
configurations rejected only because arbitrary naming conventions are
invisible. McNemar test on the 14 OPA failures transitioning to 3 yields
\(p=0.016\).

\textbf{Finding 14 (Residual oracle failures expose benchmark
artifacts).} The 3 remaining OPA failures under Rego visibility (1.6\%
of the benchmark) should not be read as a definitive model ceiling
without qualification. Further analysis shows that these failures are
benchmark artifacts: two involve self-contradictory Rego rules, and one
uses an incorrect JSON path in the OPA query. Excluding these artifacts
gives a corrected diagnostic upper bound of 176/183 (96.2\%) for GPT-4o
under the oracle setup.

\textbf{Finding 15 (Dataset design observation).} Approximately 6\% of
IaC-Eval v2 (11/186 tasks) tests implicit naming conventions rather than
infrastructure behavior. Future benchmark iterations should separate
\textbf{behavioral specifications} (resource types, attributes,
relationships) from \textbf{naming specifications} (arbitrary label
choices) to avoid conflating model capability with prompt completeness.

\subsection{Error Taxonomy: SELF\_DEFINED\_PROPERTY
Hierarchy}\label{error-taxonomy-self_defined_property-hierarchy}

\textbf{Table 8.} SELF\_DEFINED\_PROPERTY prevalence in manually coded
error analyses.

\begin{longtable}[]{@{}llccc@{}}
\toprule\noalign{}
Strategy & Model & SELF\_DEFINED errors & Total failures & Share \\
\midrule\noalign{}
\endhead
\bottomrule\noalign{}
\endlastfoot
Qwen refinement & Qwen 7B & 27 & 69 & 39.1\% \\
SIMBA & Qwen 7B & 51 & 102 & 50.0\% \\
GEPA-RAG & Qwen 7B & 20 & 87 & 23.0\% \\
GPT-4o refinement & GPT-4o & 2 & 29 & 6.9\% \\
\end{longtable}

\textbf{Finding 16 (Three-tier schema hallucination hierarchy).}
Provider-schema hallucination resolves at three distinct intervention
levels: - \textbf{Demonstration anchoring (SIMBA):} No reduction ---
50\% of failures remain SELF\_DEFINED\_PROPERTY. - \textbf{Instruction
rules (GEPA):} Partial reduction to 23\% --- abstract rules constrain
attribute generation. - \textbf{Model scaling (GPT-4o):}
Near-elimination at 6.9\% --- parametric knowledge contains correct
schemas.

This hierarchy suggests that SELF\_DEFINED\_PROPERTY is strongly tied to
model capacity and schema knowledge, not only prompt format. Instruction
optimization (GEPA) provides an intermediate-cost mitigation between
demonstration-only optimization and scaling to a stronger model.

\begin{center}\rule{0.5\linewidth}{0.5pt}\end{center}

\section{Stratified Analysis}\label{stratified-analysis}

\subsection{Performance by Difficulty
Level}\label{performance-by-difficulty-level}

\textbf{Table 9.} Pass@1 by difficulty level (selected strategies).

\begin{longtable}[]{@{}
  >{\raggedright\arraybackslash}p{(\linewidth - 12\tabcolsep) * \real{0.0909}}
  >{\centering\arraybackslash}p{(\linewidth - 12\tabcolsep) * \real{0.1515}}
  >{\centering\arraybackslash}p{(\linewidth - 12\tabcolsep) * \real{0.1515}}
  >{\centering\arraybackslash}p{(\linewidth - 12\tabcolsep) * \real{0.1515}}
  >{\centering\arraybackslash}p{(\linewidth - 12\tabcolsep) * \real{0.1515}}
  >{\centering\arraybackslash}p{(\linewidth - 12\tabcolsep) * \real{0.1515}}
  >{\centering\arraybackslash}p{(\linewidth - 12\tabcolsep) * \real{0.1515}}@{}}
\toprule\noalign{}
\begin{minipage}[b]{\linewidth}\raggedright
Strategy
\end{minipage} & \begin{minipage}[b]{\linewidth}\centering
L1 (41)
\end{minipage} & \begin{minipage}[b]{\linewidth}\centering
L2 (43)
\end{minipage} & \begin{minipage}[b]{\linewidth}\centering
L3 (52)
\end{minipage} & \begin{minipage}[b]{\linewidth}\centering
L4 (22)
\end{minipage} & \begin{minipage}[b]{\linewidth}\centering
L5 (11)
\end{minipage} & \begin{minipage}[b]{\linewidth}\centering
L6 (17)
\end{minipage} \\
\midrule\noalign{}
\endhead
\bottomrule\noalign{}
\endlastfoot
CoT Qwen 7B & 14.6\% & 27.9\% & 13.5\% & 4.5\% & 0.0\% & 0.0\% \\
Active RAG Qwen 7B & 68.3\% & 55.8\% & 50.0\% & 13.6\% & 27.3\% &
5.9\% \\
GEPA-RAG Qwen 7B & 63.4\% & 69.8\% & 57.7\% & 22.7\% & 18.2\% &
35.3\% \\
Repair GPT-4o & 87.8\% & 95.3\% & 80.8\% & 68.2\% & 63.6\% & 94.1\% \\
Rego Oracle GPT-4o & 100\% & 97.7\% & 88.5\% & 77.3\% & 90.9\% &
100\% \\
\end{longtable}

\textbf{Finding 17 (Non-monotonic difficulty).} The difficulty label is
not monotonic with observed pass rates: L6 frequently outperforms L5 for
the same strategy (e.g., Repair GPT-4o: 94.1\% on L6 vs.~63.6\% on L5).
L5 clusters around specific service families (Lex, Lightsail-database)
with deep schema constraints. Service composition matters more than the
numeric difficulty label.

\subsection{Performance by Service
Family}\label{performance-by-service-family}

\textbf{Table 10.} Pass@1 by dominant AWS service family (selected
strategies).

\begin{longtable}[]{@{}
  >{\raggedright\arraybackslash}p{(\linewidth - 14\tabcolsep) * \real{0.0789}}
  >{\centering\arraybackslash}p{(\linewidth - 14\tabcolsep) * \real{0.1316}}
  >{\centering\arraybackslash}p{(\linewidth - 14\tabcolsep) * \real{0.1316}}
  >{\centering\arraybackslash}p{(\linewidth - 14\tabcolsep) * \real{0.1316}}
  >{\centering\arraybackslash}p{(\linewidth - 14\tabcolsep) * \real{0.1316}}
  >{\centering\arraybackslash}p{(\linewidth - 14\tabcolsep) * \real{0.1316}}
  >{\centering\arraybackslash}p{(\linewidth - 14\tabcolsep) * \real{0.1316}}
  >{\centering\arraybackslash}p{(\linewidth - 14\tabcolsep) * \real{0.1316}}@{}}
\toprule\noalign{}
\begin{minipage}[b]{\linewidth}\raggedright
Strategy
\end{minipage} & \begin{minipage}[b]{\linewidth}\centering
IAM (25)
\end{minipage} & \begin{minipage}[b]{\linewidth}\centering
DynamoDB (20)
\end{minipage} & \begin{minipage}[b]{\linewidth}\centering
RDS (19)
\end{minipage} & \begin{minipage}[b]{\linewidth}\centering
S3 (15)
\end{minipage} & \begin{minipage}[b]{\linewidth}\centering
Lightsail (15)
\end{minipage} & \begin{minipage}[b]{\linewidth}\centering
ElastiCache (12)
\end{minipage} & \begin{minipage}[b]{\linewidth}\centering
Lex (8)
\end{minipage} \\
\midrule\noalign{}
\endhead
\bottomrule\noalign{}
\endlastfoot
Active RAG Qwen 7B & 24\% & 35\% & 63\% & 67\% & 60\% & 42\% & 0\% \\
GEPA-RAG Qwen 7B & 24\% & 55\% & 63\% & 67\% & 47\% & 50\% & 0\% \\
Repair GPT-4o & 56\% & 85\% & 95\% & 87\% & 87\% & 83\% & 100\% \\
\end{longtable}

\textbf{Finding 18 (IAM is the persistent weak spot).} IAM tasks have
the lowest pass rate across all strategies. Even the Rego oracle reaches
only 72\% on IAM --- the lowest family score. IAM requires constructing
precise \texttt{assume\_role\_policy} JSON embedded within HCL, forming
dependency chains that \texttt{terraform\ plan} validates strictly.

\textbf{Finding 19 (Lex is all-or-nothing).} Every Qwen strategy scores
0--12.5\% on Lex (8 tasks); GPT-4o + MCP achieves 100\%. Lex requires
nested
\texttt{intent}/\texttt{slot\_type}/\texttt{fulfillment\_activity}
blocks whose grammar is provider-v6-specific. Small models lack the
parametric knowledge; large models with retrieval can reconstruct it.

\subsection{Universally Unsolved
Tasks}\label{universally-unsolved-tasks}

Eight tasks (4.3\%) fail across all single-pass strategies evaluated:

\begin{longtable}[]{@{}
  >{\raggedright\arraybackslash}p{(\linewidth - 4\tabcolsep) * \real{0.3333}}
  >{\raggedright\arraybackslash}p{(\linewidth - 4\tabcolsep) * \real{0.3333}}
  >{\raggedright\arraybackslash}p{(\linewidth - 4\tabcolsep) * \real{0.3333}}@{}}
\toprule\noalign{}
\begin{minipage}[b]{\linewidth}\raggedright
Cluster
\end{minipage} & \begin{minipage}[b]{\linewidth}\raggedright
Tasks
\end{minipage} & \begin{minipage}[b]{\linewidth}\raggedright
Root cause
\end{minipage} \\
\midrule\noalign{}
\endhead
\bottomrule\noalign{}
\endlastfoot
Lambda zip-file PLAN\_FAIL & 4 tasks & \texttt{filebase64sha256("…zip")}
requires a real file \\
OPA capability gap & 2 tasks & True reasoning failure (or benchmark
defect) \\
Lightsail region constraint & 1 task & \texttt{availability\_zone} must
match provider region \\
Lex multi-resource composition & 1 task & Requires correct ordering of 3
dependent Lex resources \\
\end{longtable}

These 8 tasks represent the \textbf{structural lower bound} on benchmark
residual error --- unsolvable under any evaluated strategy due to
harness constraints, benchmark artifacts, or genuine model capability
gaps.

\begin{center}\rule{0.5\linewidth}{0.5pt}\end{center}

\section{Discussion}\label{discussion}

\subsection{Retrieval, Repair, and Optimization Target Different
Verifier
Stages}\label{retrieval-repair-and-optimization-target-different-verifier-stages}

The central insight of this study is that IaC generation improvements
are \textbf{stage-specific}. Table 11 summarizes the failure trajectory
for Qwen 7B across the strategy ladder. Each row accounts for the same
186 tasks: tasks first fail at \texttt{terraform\ validate}, then
surviving tasks may fail at \texttt{terraform\ plan}, then surviving
plannable tasks may fail at \texttt{opa\ eval}; the remaining tasks are
counted as PASS. The columns therefore represent a sequential verifier
funnel, not independent error categories.

\textbf{Table 11.} Failure trajectory across strategies (Qwen 7B, 186
tasks).

\begin{longtable}[]{@{}lccccc@{}}
\toprule\noalign{}
Strategy & PASS & VALIDATE\_FAIL & PLAN\_FAIL & OPA\_FAIL & pass@1 \\
\midrule\noalign{}
\endhead
\bottomrule\noalign{}
\endlastfoot
CoT & 26 & 144 & 7 & 9 & 14.0\% \\
Static RAG & 61 & 98 & 15 & 12 & 32.8\% \\
Active MCP & 79 & 82 & 10 & 15 & 42.5\% \\
Active RAG & 85 & 66 & 14 & 21 & 45.7\% \\
SIMBA & 84 & 78 & 6 & 18 & 45.2\% \\
GEPA-RAG & 99 & 57 & 11 & 19 & 53.2\% \\
Iterative refinement & 117 & 43 & 9 & 17 & 62.9\% \\
\end{longtable}

\begin{figure}[htbp]
\centering
\begin{tikzpicture}
\begin{axis}[
  xbar stacked,
  width=0.88\linewidth, height=7cm,
  bar width=10pt,
  xlabel={Task count ($n=186$)},
  symbolic y coords={CoT,{Static RAG},{Active MCP},{Active RAG},SIMBA,{GEPA-RAG},Refine},
  ytick=data,
  y dir=reverse,
  xmin=0, xmax=186,
  legend style={at={(0.5,-0.22)},anchor=north,legend columns=4,font=\footnotesize},
  tick label style={font=\footnotesize},
  label style={font=\small},
]
\addplot[fill=green!55] coordinates {(26,CoT)(61,{Static RAG})(79,{Active MCP})(85,{Active RAG})(84,SIMBA)(99,{GEPA-RAG})(117,Refine)};
\addlegendentry{PASS}
\addplot[fill=red!65] coordinates {(144,CoT)(98,{Static RAG})(82,{Active MCP})(66,{Active RAG})(78,SIMBA)(57,{GEPA-RAG})(43,Refine)};
\addlegendentry{VALIDATE\_FAIL}
\addplot[fill=orange!70] coordinates {(7,CoT)(15,{Static RAG})(10,{Active MCP})(14,{Active RAG})(6,SIMBA)(11,{GEPA-RAG})(9,Refine)};
\addlegendentry{PLAN\_FAIL}
\addplot[fill=violet!60] coordinates {(9,CoT)(12,{Static RAG})(15,{Active MCP})(21,{Active RAG})(18,SIMBA)(19,{GEPA-RAG})(17,Refine)};
\addlegendentry{OPA\_FAIL}
\end{axis}
\end{tikzpicture}
\caption{Failure-stage decomposition across Qwen~7B strategies. Improvements primarily reduce \textsc{validate\_fail}; OPA failures grow modestly as configurations progress further through the pipeline (failure promotion).}
\label{fig:failure-stages}
\end{figure}
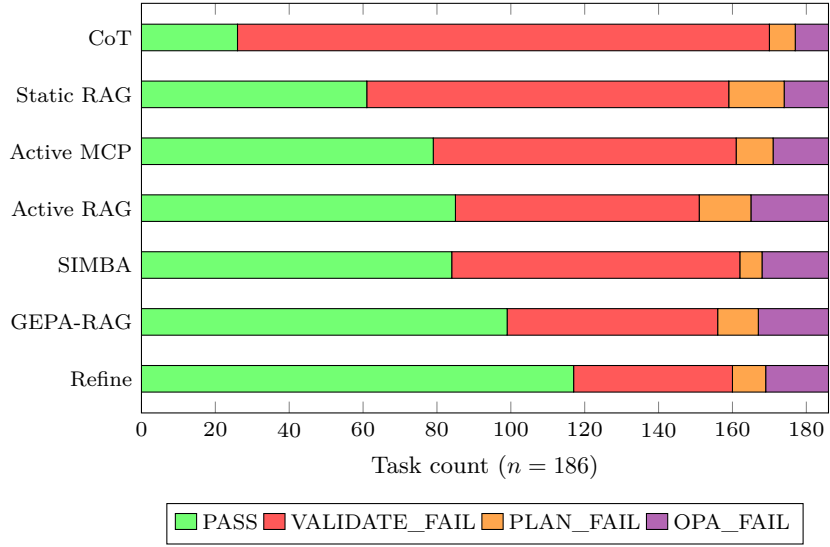

\textbf{Figure 3.} Failure-stage decomposition across Qwen strategies.
The figure uses the same columns as Table 11: PASS, VALIDATE\_FAIL,
PLAN\_FAIL, and OPA\_FAIL. The ``Iterative refinement'' row is the same
strategy called ``+ Repair'' in Table 4: the model generates an initial
answer, receives verifier errors, and retries up to three times. In this
section we use ``iterative refinement'' for the strategy name and
``repair'' for the retry mechanism. The figure should be read
left-to-right as a movement through the verifier funnel. The largest
visible effect of retrieval, GEPA, and iterative refinement is the
shrinking VALIDATE\_FAIL segment: Qwen moves from 144 validation
failures under CoT to 43 after refinement. PLAN\_FAIL remains
comparatively small and stable. OPA\_FAIL does not disappear because
more generated configurations now survive validation and planning,
meaning the policy gate receives more candidates to evaluate.

Three patterns emerge: 1. \textbf{Most gains come from reducing schema
failures}: VALIDATE\_FAIL falls from 144 to 43 ($-$70\%), showing that
retrieval, optimization, and repair mainly improve provider-schema
correctness. 2. \textbf{PLAN\_FAIL is approximately constant} (6--15
range): This is the Lambda zip-file cluster --- an immobile harness
artifact. 3. \textbf{OPA failures are promoted failures, not necessarily
regressions}: OPA grows from 9 to 17 because more tasks reach the policy
gate. Once validation errors are reduced, prompt or policy enrichment
becomes necessary to address these later-stage failures.

\subsection{GEPA and RAG Are
Complementary}\label{gepa-and-rag-are-complementary}

GEPA optimizes \emph{reasoning strategy} (how to generate Terraform from
retrieved docs); RAG provides \emph{knowledge content} (which docs to
retrieve). The +7.5 pp GEPA improvement comes from filling gaps that RAG
cannot address: API constraint knowledge that appears in AWS validation
errors, not in provider documentation. This defines a
\textbf{complementarity principle}: use RAG for schema knowledge gaps;
use GEPA for implicit constraint gaps.

\subsection{The Pareto Frontier of Accuracy
vs.~Latency}\label{the-pareto-frontier-of-accuracy-vs.-latency}

\textbf{Table 12.} Pareto frontier --- accuracy vs.~latency (all
strategies).

\begin{longtable}[]{@{}
  >{\raggedright\arraybackslash}p{(\linewidth - 8\tabcolsep) * \real{0.1579}}
  >{\raggedright\arraybackslash}p{(\linewidth - 8\tabcolsep) * \real{0.1579}}
  >{\centering\arraybackslash}p{(\linewidth - 8\tabcolsep) * \real{0.2632}}
  >{\centering\arraybackslash}p{(\linewidth - 8\tabcolsep) * \real{0.2632}}
  >{\raggedright\arraybackslash}p{(\linewidth - 8\tabcolsep) * \real{0.1579}}@{}}
\toprule\noalign{}
\begin{minipage}[b]{\linewidth}\raggedright
Tier
\end{minipage} & \begin{minipage}[b]{\linewidth}\raggedright
Strategy
\end{minipage} & \begin{minipage}[b]{\linewidth}\centering
pass@1
\end{minipage} & \begin{minipage}[b]{\linewidth}\centering
Latency
\end{minipage} & \begin{minipage}[b]{\linewidth}\raggedright
Infrastructure
\end{minipage} \\
\midrule\noalign{}
\endhead
\bottomrule\noalign{}
\endlastfoot
Fast single-pass & CoT Qwen 7B & 14.0\% & 42s & None \\
Retrieval-augmented & Active RAG & 45.7\% & 36s & ChromaDB \\
Optimized (teacher-free) & SIMBA & 45.2\% & 48s & None (one-time 2h
train) \\
Optimized (teacher-backed) & GEPA-RAG & 53.2\% & 40s & ChromaDB +
GPT-5.4 (compile only) \\
Iterative (local) & Repair Qwen 7B & 62.9\% & 203s & MCP + verifier \\
Iterative (frontier) & Repair GPT-4o & 84.4\% & 98s & MCP + verifier +
API cost \\
Oracle & Rego injection & 93.0\% & 86s & + Rego policy visibility \\
\end{longtable}

For production deployment: - \textbf{Latency-critical (\textless=50s):}
GEPA-RAG (53.2\%) is Pareto-optimal in the fast tier. -
\textbf{Accuracy-critical:} GPT-4o + Repair (84.4\%) at 4× latency cost.
- \textbf{Infrastructure-constrained:} SIMBA (45.2\%) requires no
runtime dependencies beyond the model.

\subsection{Implications for IaC Generation
Research}\label{implications-for-iac-generation-research}

\begin{enumerate}
\def\labelenumi{\arabic{enumi}.}
\tightlist
\item
  \textbf{Evaluation should preserve verifier-stage information.} A
  single pass@1 cannot distinguish schema hallucination from policy
  non-compliance. Stage-decomposed reporting is essential.
\item
  \textbf{Retrieval and repair are complementary, not interchangeable.}
  Retrieval prevents errors before generation; repair corrects them
  after. They address different failure classes.
\item
  \textbf{Policy compliance requires different information than schema
  correctness.} The Rego oracle shows that 79\% of policy failures arise
  because the prompt omits constraints the policy enforces --- not
  because the model cannot generate compliant code.
\item
  \textbf{Instruction optimization (GEPA) is automated system-prompt
  engineering} and is more effective than demonstration optimization
  (SIMBA) for schema-constrained domains.
\item
  \textbf{Schema hallucination is primarily a model-scale limitation}
  addressable by instruction rules (partial) or model scaling (full),
  not by demonstrations or retrieval alone.
\end{enumerate}

\begin{center}\rule{0.5\linewidth}{0.5pt}\end{center}

\section{Observability and
Reproducibility}\label{observability-and-reproducibility-1}

\subsection{Experimental
Infrastructure}\label{experimental-infrastructure}

All experiments run on a Kubernetes cluster provisioned via Terraform. The observability stack comprises:
- \textbf{MLflow v3}: Persistent experiment tracking with per-span
tracing of every LM call. One experiment per strategy (14 experiments,
\textasciitilde3,400 total traces). - \textbf{Phoenix (Arize)}:
Real-time trace visualization via OpenInference protocol. - \textbf{DSPy
autolog}: Automatic instrumentation of all DSPy module calls, predictions, agent steps, and LM invocations.

\subsection{Trace-Level Verifiability}\label{trace-level-verifiability}

Every numerical result in this paper is traceable to a specific experiment, a per-task execution trace, and individual span records capturing model inputs, outputs, tool calls, and timing. This level of reproducibility exceeds the standard for code-generation benchmarks, where typically only aggregate metrics are reported. Our methodology enables independent researchers to verify any individual task outcome by examining its trace, reproduce error taxonomy results by parsing span-level outputs, and confirm statistical claims from the released per-task pass/fail vectors.

\subsection{Artifact Availability}\label{artifact-availability}

\begin{longtable}[]{@{}
  >{\raggedright\arraybackslash}p{(\linewidth - 4\tabcolsep) * \real{0.3333}}
  >{\raggedright\arraybackslash}p{(\linewidth - 4\tabcolsep) * \real{0.3333}}
  >{\raggedright\arraybackslash}p{(\linewidth - 4\tabcolsep) * \real{0.3333}}@{}}
\toprule\noalign{}
\begin{minipage}[b]{\linewidth}\raggedright
Artifact
\end{minipage} & \begin{minipage}[b]{\linewidth}\raggedright
Location
\end{minipage} & \begin{minipage}[b]{\linewidth}\raggedright
Description
\end{minipage} \\
\midrule\noalign{}
\endhead
\bottomrule\noalign{}
\endlastfoot
IaC-Eval v2 dataset & \texttt{huggingface.co/datasets/iac-eval-v2/iac-eval-v2} & 186 tasks with natural-language prompts, Rego intent specifications, and reference HCL solutions \\
Per-task result files & Released with the paper & Per-strategy JSON files with pass/fail outcome, verifier-stage classification, attempt count, and elapsed time for each of the 186 tasks \\
Compiled optimizer states & Released with the paper & Serialized GEPA and SIMBA optimizer states encoding learned instructions and demonstrations \\
Trace query toolkit & Released with the paper & Scripts for programmatic access to all MLflow traces via the REST API \\
Infrastructure code & Released with the paper & Full Terraform configuration for the Kubernetes-based observability and inference platform \\
\end{longtable}

\begin{center}\rule{0.5\linewidth}{0.5pt}\end{center}

\section{Threats to Validity}\label{threats-to-validity}

\textbf{Internal validity.} All metrics are single-run at
temperature=0.0. This makes the evaluation deterministic for a fixed
model endpoint, but it does not measure run-to-run variance from
hosted-model changes, inference nondeterminism, or backend routing.
Paired McNemar tests reduce noise for strategy comparisons, but
borderline results such as GEPA vs.~SIMBA (\(p=0.059\)) should be
treated as suggestive.

\textbf{Construct validity.} \texttt{terraform\ validate},
\texttt{terraform\ plan}, and \texttt{opa\ eval} are strong proxies for
deployable and policy-compliant Terraform, but they are not a complete
production deployment test. Some PLAN\_FAIL cases, especially Lambda
zip-file tasks, reflect benchmark-harness constraints rather than pure
model errors.

\textbf{External validity.} IaC-Eval v2 covers AWS/Terraform only. The
failure taxonomy is likely relevant to other declarative IaC languages,
but generalization to Azure/Bicep, GCP, Pulumi, or Kubernetes manifests
requires separate evaluation.

\textbf{Optimizer validity.} GEPA uses a moderate optimization budget, while SIMBA is run at a deliberately minimal configuration to establish a lower bound. The optimizer comparison therefore supports conclusions about the evaluated configurations, not final ceilings for either optimizer.

\textbf{Policy leakage.} The Rego-injection experiment is a diagnostic
oracle, not a realistic production strategy. It measures how many OPA
failures are caused by missing policy information in the prompt, but it
should not be compared directly to non-oracle systems as a deployable
method.

\textbf{Cost measurement.} LiteLLM did not consistently forward per-call
token accounting for all local and proxied runs. We therefore report
latency and pass rates as primary metrics and treat cost estimates as
secondary.

\begin{center}\rule{0.5\linewidth}{0.5pt}\end{center}

\section{Future Work}\label{future-work}

\begin{enumerate}
\def\labelenumi{\arabic{enumi}.}
\tightlist
\item
  \textbf{GEPA + Repair composition.} Combine GEPA-optimized
  instructions (53.2\% first-pass) with iterative refinement to test
  whether compile-time and runtime corrections compose cleanly.
\item
  \textbf{Scaled SIMBA evaluation.} Run SIMBA at steps=8,
  num\_candidates=8 (\textasciitilde30h) to determine whether
  teacher-free optimization can match GEPA without a frontier model.
\item
  \textbf{Adaptive optimizer selection.} Route tasks to SIMBA
  (demo-based) or GEPA (rule-based) based on predicted failure mode,
  analogous to Adaptive-RAG for retrieval strategy.
\item
  \textbf{Multi-provider generalization.} Evaluate whether the
  verifier-stage methodology and failure taxonomy transfer to
  Azure/Bicep, GCP/Deployment Manager, and Pulumi.
\item
  \textbf{Policy-aware prompt enrichment.} The Rego oracle shows policy
  visibility resolves 79\% of OPA failures. Developing practical,
  privacy-preserving policy summarization for production use is a
  natural next step.
\end{enumerate}

\begin{center}\rule{0.5\linewidth}{0.5pt}\end{center}

\section{Conclusion}\label{conclusion}

This paper demonstrates that reliable IaC generation requires
\textbf{stage-aware evaluation and intervention design}. On IaC-Eval v2
(186 Terraform/AWS tasks), we show that:

\begin{itemize}
\tightlist
\item
  \textbf{Retrieval} (ReAct + RAG) reduces provider-schema failures from
  144 to 66 tasks, raising pass@1 from 14.0\% to 45.7\% (\(p<0.0001\)).
\item
  \textbf{GEPA instruction optimization} adds +7.5 pp (\(p=0.026\)) by
  encoding verifier-derived constraint rules into instruction text,
  showing that prompt optimization can improve verifiable IaC
  generation.
\item
  \textbf{Iterative refinement} recovers validation and planning
  failures to reach 84.4\% (GPT-4o), but cannot address policy failures
  without policy visibility.
\item
  \textbf{Schema hallucination} (SELF\_DEFINED\_PROPERTY) is the
  dominant failure mode for local models, addressable only partially by
  instruction rules and fully by model scaling.
\item
  \textbf{79\% of residual policy failures} are information gaps in the
  prompt--policy interface, not model capability failures.
\end{itemize}

The broader lesson is that a single pass@1 score hides the structure of
IaC generation failures. By reporting per-stage metrics, applying paired
statistical tests, and maintaining trace-level reproducibility, we
provide an empirical foundation for designing IaC agents that target
specific failure classes rather than optimizing an aggregate score.

\begin{center}\rule{0.5\linewidth}{0.5pt}\end{center}

\section*{References}\label{references}
\addcontentsline{toc}{section}{References}

{[}Agrawal et al., 2026{]} Agrawal, L. A. et al.~``GEPA: Reflective
Prompt Evolution Can Outperform Reinforcement Learning.'' ICLR 2026
(Oral). arXiv:2507.19457.

{[}Anthropic et al., 2024{]} Anthropic et al.~``Model Context
Protocol.'' Specification, November 2024.
https://modelcontextprotocol.io

{[}Chen et al., 2021{]} Chen, M. et al.~``Evaluating Large Language
Models Trained on Code.'' arXiv:2107.03374.

{[}HashiCorp, 2024{]} HashiCorp. ``Terraform Documentation.''
https://developer.hashicorp.com/terraform/docs

{[}Jeong et al., 2024{]} Jeong, S. et al.~``Adaptive-RAG: Learning to
Adapt Retrieval-Augmented Large Language Models through Question
Complexity.'' NAACL 2024. arXiv:2403.14403.

{[}Khattab et al., 2024{]} Khattab, O. et al.~``DSPy: Compiling
Declarative Language Model Calls into Self-Improving Pipelines.'' ICLR
2024. arXiv:2310.03714.

{[}Kon et al., 2024{]} Kon, P. T. J. et al.~``IaC-Eval: A Code
Generation Benchmark for Cloud Infrastructure-as-Code Programs.''
NeurIPS 2024 (Datasets \& Benchmarks). DOI:10.52202/079017-4273.

{[}Kon et al., 2025{]} Kon, P. T. J. et al.~``EXP-Bench: Can AI Conduct
AI Research Experiments?'' arXiv:2505.24785.

{[}Lewis et al., 2020{]} Lewis, P. et al.~``Retrieval-Augmented
Generation for Knowledge-Intensive NLP Tasks.'' NeurIPS 2020.
arXiv:2005.11401.

{[}Madaan et al., 2023{]} Madaan, A. et al.~``Self-Refine: Iterative
Refinement with Self-Feedback.'' NeurIPS 2023. arXiv:2303.17651.

{[}Nashid et al., 2023{]} Nashid, N. et al.~``Retrieval-Based Prompt
Selection for Code-Related Few-Shot Learning.'' ICSE 2023.

{[}Olausson et al., 2023{]} Olausson, T. X. et al.~``Is Self-Repair a
Silver Bullet for Code Generation?'' arXiv:2306.09896.

{[}Opsahl-Ong et al., 2024{]} Opsahl-Ong, K. et al.~``Optimizing
Instructions and Demonstrations for Multi-Stage Language Model
Programs.'' arXiv:2406.11695.

{[}Yao et al., 2023{]} Yao, S. et al.~``ReAct: Synergizing Reasoning and
Acting in Language Models.'' ICLR 2023. arXiv:2210.03629.

{[}Zhang et al., 2025{]} Zhang, T. et al.~``Deployability-Centric
Infrastructure-as-Code Generation.'' FSE 2025. arXiv:2506.05623.

\begin{center}\rule{0.5\linewidth}{0.5pt}\end{center}

\appendix

\section{Dataset Composition}\label{dataset-composition}

\begin{longtable}[]{@{}llcc@{}}
\toprule\noalign{}
Dimension & Bucket & Tasks & Share \\
\midrule\noalign{}
\endhead
\bottomrule\noalign{}
\endlastfoot
Difficulty & L1 & 41 & 22.0\% \\
& L2 & 43 & 23.1\% \\
& L3 & 52 & 28.0\% \\
& L4 & 22 & 11.8\% \\
& L5 & 11 & 5.9\% \\
& L6 & 17 & 9.1\% \\
Resource count & 1 & 84 & 45.2\% \\
& 2--3 & 51 & 27.4\% \\
& 4--6 & 45 & 24.2\% \\
& 7+ & 6 & 3.2\% \\
\end{longtable}

\section{GEPA Evolved Instruction
(Excerpt)}\label{gepa-evolved-instruction-excerpt}

The GEPA-optimized instruction
evolved from a 3-word default (``Generate Terraform HCL'') to a 40+ line
domain specification. Key rules learned:

\begin{verbatim}
- aws_elasticache_user: passwords must be 16–128 chars, use random_password
- aws_elasticache_user: engine must be exactly "redis" (lowercase)
- aws_dynamodb_contributor_insights: requires BOTH aws_dynamodb_table 
  AND aws_dynamodb_contributor_insights resources
- aws_msk_cluster: Prometheus monitoring under open_monitoring.prometheus,
  NOT under broker_node_group_info
- aws_lightsail_database: requires blueprint_id = "postgres_12"
- aws_rds_cluster: Aurora MySQL uses engine = "aurora-mysql", NOT "aurora"
\end{verbatim}

\section{SIMBA Selected
Demonstration}\label{simba-selected-demonstration}

SIMBA selected Task 265 (basic \texttt{aws\_iam\_group} with a single
\texttt{name} attribute) as the demonstration for all three pipeline
modules. This reflects SIMBA's contrast-maximization criterion:
trivially simple tasks produce the largest score-variance gap between
good and bad trajectories, making them maximally informative under
SIMBA's optimization objective --- but minimally informative for complex
schema-constrained generation.

\section{Statistical Methods}\label{statistical-methods}

\textbf{McNemar's test} (continuity-corrected): For paired binary
outcomes \((a_{ij})\) where \(b\) = tasks strategy A passes and B fails,
\(c\) = tasks B passes and A fails:

\[\chi^2 = \frac{(|b - c| - 1)^2}{b + c}\]

with 1 degree of freedom. Reported \(p\)-values are two-tailed.

\textbf{Wilson score interval}: For \(k\) successes in \(n\) trials at
confidence level \(z_{\alpha/2}\):

\[\tilde{p} = \frac{k + z^2/2}{n + z^2}, \quad w = \frac{z}{n + z^2}\sqrt{kq/n + z^2/4n}\]

where \(q = n - k\). The interval is \([\tilde{p} - w, \tilde{p} + w]\).

\section{Artifact Availability and
Reproducibility}\label{artifact-availability-and-reproducibility}

The released artifact package includes:

\begin{itemize}
\tightlist
\item
  \textbf{Dataset:} IaC-Eval v2 on Hugging Face (\texttt{huggingface.co/datasets/iac-eval-v2/iac-eval-v2}) --- 186 tasks with prompts, Rego specifications, and reference solutions.
\item
  \textbf{Per-task results:} One result file per strategy, containing per-task pass/fail outcome, verifier-stage classification, attempt count, and elapsed time.
\item
  \textbf{Compiled optimizer states:} Serialized GEPA and SIMBA pipeline states encoding learned instructions and demonstrations.
\item
  \textbf{MLflow trace archive:} All experiments and task-level execution traces accessible via the MLflow REST API.
\item
  \textbf{Statistical analysis:} McNemar tests and Wilson confidence intervals computed from per-task pass/fail vectors, with the computation script included.
\item
  \textbf{Infrastructure code:} Full Terraform configuration for the Kubernetes-based platform.
\end{itemize}

\end{document}